# Defense against SYN-Flood Denial of Service Attacks Based on Learning Automata

Masoud Bekravi*[1], Shahram Jamali[2] and Gholam Shaker[3]

*[1] Department of Computer Engineering, Ardabil Branch, Islamic Azad University, Ardabil, Iran

[2] Department of Computer Engineering, University of Mohaghegh Ardabili, Ardabil, Iran

[3] Young Researchers Club, Ardabil Branch, Islamic Azad University, Ardabil, Iran

**Abstract**

SYN-flooding attack uses the weakness available in TCP's three-way handshake process to keep it from handling legitimate requests. This attack causes the victim host to populate its backlog queue with forged TCP connections. In other words it increases Ploss (probability of loss) and Pa (buffer occupancy percentage of attack requests) and decreases Pr (buffer occupancy percentage of regular requests) in the victim host and results to decreased performance of the host. This paper proposes a self-managing approach, in which the host defends against SYN-flooding attack by dynamically tuning of its own two parameters, that is, m (maximum number of half-open connections) and h (hold time for each half-open connection). In this way, it formulates the defense problem to an optimization problem and then employs the learning automata (LA) algorithm to solve it. The simulation results show that the proposed defense strategy improves performance of the under attack system in terms of Ploss, Pa and Pr.

## 1. Introduction

Security has become necessary in a world where more services are relying on internet technology. For this reason, it has attracted a lot of attention in various areas of communication networks [16, 25, and 9]. One of the security breaches is denial-of-service (DoS) attack. A DoS attack can be considered as an attempt of attackers to prevent legal users from gaining a normal network service [27, 30, and 14]. Recent evaluations [15, 18] show that DoS attacks ranks at the fourth place in the list of the most important attack classes for information systems. More than 90% of distributed denial-of-service (DDoS) attacks exploit a system's transmission control protocol (TCP) [12]. A well-known DoS attack is SYN-flooding attack. A TCP connection is established in what is known as a 3-way handshake. When a client efforts to start a TCP connection to a server, firstly, the client requests a connection by sending a SYN packet to the server. Then, the server returns a SYN-ACK, to the client. Finally, the client acknowledges the SYN-ACK with an ACK, at which point the connection is established and data transfer commences [11, 6]. In a SYN flooding attack, attackers use this protocol to their benefit. The attacker sends a large number of SYN packets to the server. Each of these packets has to be handled like a connection request by the server, so the server must answer with a SYN-ACK. The attacker does not answer to the SYN-ACK, which will cause the server to have a half-open connection. The result is that the server is left waiting for a reply from a large quantity of connections. There are a limited number of connections a server can handle. Once all of these are in use, waiting for connections that will never come, no new connections can be made whether valid or not. There are some proposed defenses for this attack. Zuquete [3] proposes SYN cookies to defend against SYN-flooding attacks. A SYN-flood detection approach was proposed in [13]. This approach monitors the difference between the number of SYN segments and the number of FIN or RST segments since, under normal TCP behavior, each SYN will correspond to a FIN or RST. Therefore, a sharp rise in difference between the number of SYNs and FINs/RSTs, within a certain time frame, is indicative of a SYN flooding attack. Chang [23] mentioned a simple queuing model for the SYN-flooding attack. Long [19] proposed two queuing models for the DoS attacks in order to obtain the packet delay jitter and the loss probability. Peng [26] compiled an IP address database of previous successful connections. When a network was suffering from traffic congestion, an IP address that did not appear in the database was construed as more suspicious. As another work [6] proposes an autonomous approach in which the victim host defends against SYN-flooding attack by itself and does not involves ISP, router and other network devices. Ling [28] proposed a defense procedure that uses the edge routers that connect end hosts to the Internet to store and detect whether the outgoing SYN, ACK or incoming SYN/ACK segment is valid. This is accomplished by maintaining a mapping table of the outgoing SYN segments and incoming SYN/ACK segments, and creating the destination and source IP address database. In Ming [29] a probabilistic drop scheme is given for implementation in a host server to mitigate SYN-flooding attacks. It proposes an analysis for this scheme, and a general principle for evaluation of the probability of successful connection establishment is presented.

We believe that to face the problem of SYN-flooding, there is a need for algorithm which is independent and is aware of the dynamic traffic of the network and changes the defense parameters of the system according to network traffic conditions. The parameters noted in this paper are the maximum number of half-open connections (m) and the hold time (h) of these connections where by the use of learning automata algorithm, the optimized values of these parameters are determined with respect to network conditions.

First we will model the system under attack by use of queuing theory, and then we will discuss the learning automata algorithm, and using this algorithm we will offer our methodology, and the way to mapping the learning automata algorithm and will address the issue of SYN-flooding attacks.

## 2. Learning Automata

Learning automata is an abstract model which randomly selects one action out of its finite set of actions and performs it on a random environment. Environment then evaluates the selected action and responds to the automata with a Reinforcement signal. Based on selected action, and received signal, the automata updates its internal state and selects its next action. Fig. 1 depicts the relationship between an automaton and its environment [20].

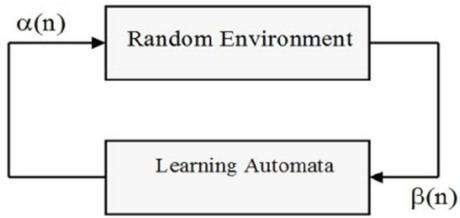

Fig. 1 Relationship between learning automata and its environment

Environment can be defined by the triple $E = \{\alpha, \beta, c\}$
where $\alpha = \{\alpha_1, \alpha_2, \ldots, \alpha_r\}$ represents a finite input set, $\beta = \{\beta_1, \beta_2, \ldots, \beta_r\}$ represents the output set, and $c = \{c_1, c_2, \ldots, c_r\}$ is a set of penalty probabilities, where each element $c_i$ of c corresponds to one input of action $\alpha_i$. An environment in which β can take only binary values 0 or 1 is referred to as P-model environment. A further generalization of the environment allows finite output sets with more than two elements that take values in the interval [0, 1]. Such an environment is referred to as Q-model. Finally, when the output of the environment is a continuous random variable which assumes values in the interval [0, 1], it is referred to as an S-model.

Learning automata are classified into fixed-structure stochastic, and variable-structure stochastic. In the following, we consider only variable-structure automata.

A variable-structure automaton is defined by the quadruple $LA = \{\alpha, \beta, p, T\}$ in which $\alpha = \{\alpha_1, \alpha_2, \ldots, \alpha_r\}$ represents the action set of the automata, $\beta = \{\beta_1, \beta_2, \ldots, \beta_r\}$ represents the input set, $p = \{p_1, p_2, \ldots, p_r\}$ represents the action probability set, and finally $P(n+1) = T[\alpha(n), \beta(n), p(n)]$ represents the learning algorithm.

This automaton operates as follows. Based on the action probability set, automaton randomly selects an action $\alpha_i$ and performs it on the environment. After receiving the environment's reinforcement signal, automaton updates its action probability set based on Eq. 1 for favorable responses, and Eq. 2 for unfavorable ones.

$$p_i(n+1) = p_i(n) + a.(1 - p_i(n)) \qquad (1)$$
$$p_i(n+1) = p_i(n) - a.p_i(n) \quad , \quad \forall j \quad j \neq i$$

$$p_i(n+1) = (1-b).p_i(n) \qquad (2)$$

$$p_i(n+1) = \frac{b}{r-1} + (-b).p_i(n) \quad , \quad \forall j \quad j \neq i$$

Learning automata is a stochastic model operating in the framework of reinforcement learning [24]. Reinforcement learning or learning with a critic is a framework of learning problems in which the teacher or the environment does not indicate the correct action, but provides only a scalar evaluative response to the selection of an action by the learner. Learning automata can be classified under the reinforcement learning schemes in the category of temporal-difference (TD) learning methods. TD learning is a combination of Monte Carlo ideas and dynamic programming ideas. Like Monte Carlo methods, TD methods can learn directly from raw experience without a model of the environment's dynamics. Like dynamic programming, TD methods update estimates based in part on other learned estimates, without waiting for a final outcome [24]. Q-learning [7, 10], Actor-Critic methods [1] and R-learning [2] are other samples of TD methods. Learning automata differs from other TD methods in the following two ways; 1.The representation of the internal states (a set of action probabilities) and 2. The updating method of the internal states (Eq.1 and Eq.2).

Learning automata has found applications in parameter optimization, statistical decision making, telephone routing, pattern recognition, game playing, natural language processing, modelling biological learning systems and object partitioning [4]. Furthermore, learning automata is proved to perform well in the dynamic environments of computer networks. It is used in wireless networks for adaptive rate control [17], bandwidth control [21] and designing reliable transport layer protocol (Learning-TCP) [5].

## 3. Queuing model

Since queues provide the most intuitive language for explaining traffic and its dependence structure, in the network environment [8], in this work we use queuing theory to draw a defense map against SYN- flooding attacks. Although a computer system includes several resources, for simplicity, we consider only one resource that is, memory and corresponding backlog buffer. In this model, all connection requests share the same backlog buffer. When a request arrives at the system, the system instantly receives a buffer space of the backlog queue upon finding an inactive buffer space and is blocked otherwise. Now, consider a server under the SYN-flooding attacks. Assume that in this computer each half-open connection is held for at most a period of time h and at most m concurrent half-open connections are allowed. We assume that a half-open connection for a regular request packet is held for a chance time which is exponentially distributed with parameter µ The arrivals of the regular request packets and the attack packets are both Poisson processes with rates $\lambda 1$ and $\lambda 2$ respectively. The two arrival processes are independent of each other and of the holding times for half-open connections. Obviously, when the system is under attack, then number of pending connections increases and in a point in which there is no more room for pending connection to be saved the arriving packets will be blocked. This leads to increased number of lost connections. On the other hand, when a server is under SYN-flooding attacks, half-open connections can quickly consume all the memory allocated for the pending connections and prevent the victim from further accepting new requests, leading to the well-known buffer overflow problem. In this case, less percentage of buffer space is occupied by legal requests, and major part of this space will be allocated to the attacker requests.

## 4. Discussion

As said earlier, as the requests with SYN packets enter the server, the TCP protocol places them in the backup buffer, and allocates necessary resources from the backup buffer for the establishment of a complete connection. This state is called the half-open state.
On the other hand, the number of half-open confections that a server can create one limited and have a maximum value. Also the holding time of these half-open confections is a fixed constant our proposed algorithm is to change these two parameters, to dynamic condition with respect to the network conditions. For this reason, we define the following parameters.
Ploss: a package is closed when, due to buffer being closed, cannot respond to a received request to create a connection. There for we define the Ploss as the ratio of the total number blocked packages to the total numbers of packages that have entered the server.
Percent owner-ship of buffer via the normal requests (RRROP): When a half-open connection is created, it is placed in the buffer. The average ratio of the number of half-open connections created by normal request, to the maximum number of half-open connections that the server can create is called percent ownership of buffer by the normal request.
Percent ownership of buffer by the attack requests (Pa): The average ratio of number of half-open connections created by attack requests to the maximum number of half-open connection that can be created by the server is called percent ownership of buffer via attack requests.
In order to increase the capability of a server to providing services, the value of Ploss must be sufficiently small. Also in order that a server provides more services to the normal requests, the value of buffer ownership by normal requests must be sufficiently big, and the time of ownership of buffer be the attack requests must be sufficiently small.
Therefore objectives of this paper are:
Reducing the value of request blockage.
Increase of percent and time of occupancy of buffer by normal requests
Reduction of percent and time of buffer occupancy by attack requests
We use this information and define the objective function for LA algorithm as:

$$\text{Maximize} \frac{Pr}{Pa*Ploss} \quad (3)$$

Therefor the higher the value of this function, the higher the ability of server to provide services. By use of this objective function we warp to use the LA algorithm to increase the functionality of the server that executes this very algorithm. Noting the common factors between the SYN-flooding attack and the learning automata, the LA algorithm is defined following. The server utilizes the initial (h, m) Parameter and initiates providing services and at each round of modelling investigates the objective Function. If the new objective function is bigger than the best objective Function and better than the previous one , then the previous values of (h, m) is used, otherwise the (h, m) value is updated the algorithm continues the execution until it reaches the desired stop Condition.
The stages of the proposed algorithm are:
Start
*1- Initialization of variables.*
*2- In each round of simulation.*
*3-1- calculation of objective function.*

*3-2- comparison with best previous objective function and finding a bigger objective function (use of automata memory)*
*3-3-Finding of designed (h, m) that result in maximized objective function (use of learned information)*
*4-continuation of service processes.*
*5-end*

### 4.1 Definition of accidental automata in SYN-flooding

The TCP protocol in process of establishing connections uses the two main parameters (h, m) to control the holding-time of half-open connections and also the number of these connections. Therefore the two parameters (h, m) influence the behavior of TCP protocol. In the TCP protocol the values of (h, m) are constant. In this paper the (h, m) parameters via use of learning automata algorithm are changed dynamically, and by moving to an optimal point, cause the early omission of the half open connections, allocated for attack requests. Also the (h, m) parameters make the total number of the system connections a function of the degree of attack and cause operational improvement of TCP in challenging the SYN-flood attacks.

### 4.2 Suitability

One of the main specification of any algorithm is its suitability. Our measures for determining suitability of a proposed algorithm are Ploss, Pr, and Pa. In various steps of execution of an algorithm the closer these measures one to the optimal values, the better the operation of the proposed algorithm. Fig. 2 shows the evaluation process of the LA algorithm:

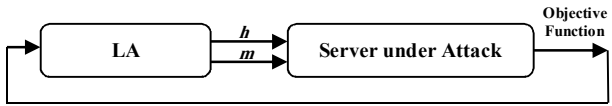

Fig. 2 Calculation of suitability and tuning the values of *(h, m)*

### 4.3 The procedure of mapping the LA method and defense in facing SYN-flooding Attacks

In this problem the victim's memory buffer includes a capacity of m with busy-time of $h_1, h_2, ..., h_m$ and there are n requests, with ith request with a value of $p_i$ and keeps the buffer occupied for a time-interval $h_i$, with a remaining buffer capacity of $w_i$. The goal is to provide services to a subset of the normal requests, deletion of the attack requests, blocking the attack requests, and have high efficacy of service. A request is either a normal request, or an attack request, the problem is defined as following:

$$\text{minimize} \sum_{i=1}^{n}(p_i \cdot x_i) \quad (4)$$
$$\text{subject to } M_i : \sum_{i=1}^{n}(w_i \cdot x_i) < h_i, \forall\, i \in 1 \dots c$$
$$x_i \in \{0,1\}, \forall\, i \in 1 \dots n \quad (5)$$

Where xi is the decision variable with respect to waiting ith request. If the ithe request is a normal request, then $x_i$ is 1, else $x_i$ is zero. The variables $w_i$, $p_i$, $h_i$ cannot assume negative values.

In the proposed algorithm the problem is modeled by a complete graph, where every node of the graph is modeled with respect to a request in the defense problem. Each node of the graph is equipped with two operations, choosing of the request to be placed in the buffer, or the deletion of the request from the buffer.
Fig. 3 shows the Flowchart of work of proposal method:

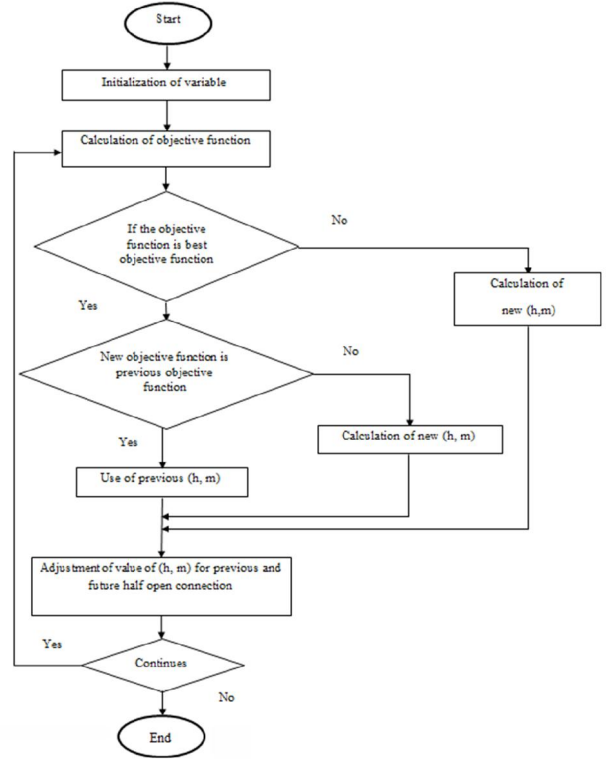

Fig. 3 Flowchart of work of proposal method

All these processes are executed repeatedly until an optimal solution is reached, or we reach the condition for maximum required repetition of the algorithm.

## 5. Result and Simulation

Here, we study the proposed defense scheme by using two essential security metrics namely, the Probability of Success of Attack and the Buffer Utilization Efficiency. These two metrics represent how severe the SYN-flooding attacks affect the system performance. For this purpose, we follow manner of [30] and give some numerical examples to exhibit how to quantify these security metrics. Let λ1=10 as the parameter for the Poisson arrival process of the regular request packets and λ2=kλ1, as the Poisson arrival process of the attack request packets, in which, k represent the ratio between arrival rates of the attack packets and the regular request packets. We use the exponential distribution with the parameter μ=100/s as the service time of regular request packets, and it could represent the strictness of congestions in the network. In order to study proposed approach performance in wide range of attack intensity, we change k from 0 to 2. Total number of connection request is considered 50000 requests, some of them are legal requests and others are attack connection requests. As a reference point we compare our approach with Linux in which m=128 and h=75s statically [22].

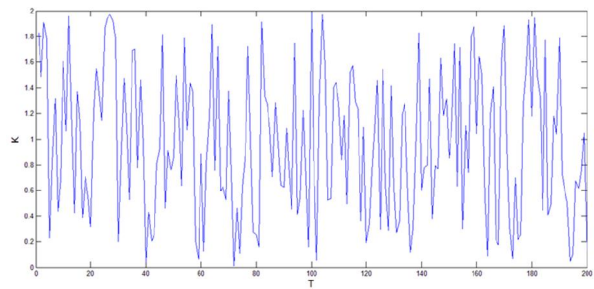

Fig. 4 Attack intensity applied to the under attack server (*k*)

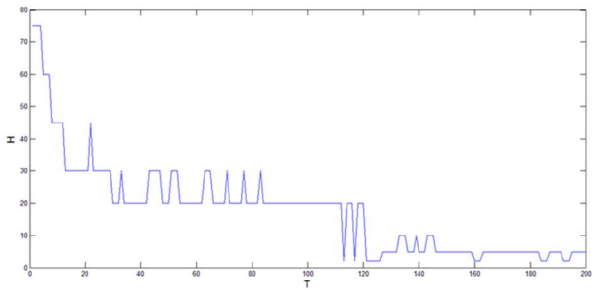

Fig. 5 *h* changes

Fig. 4, 5 respectively show the tendency of change in k and h.

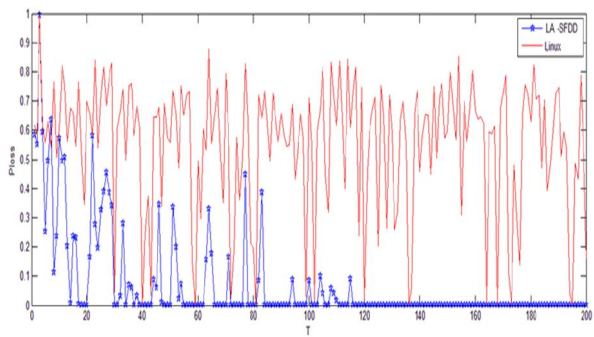

Fig. 6 probability of loss

Fig. 6 shows the probability of blockage of requests. The graph shows that the values of the blocked requests of the LA defense algorithm are much lower.

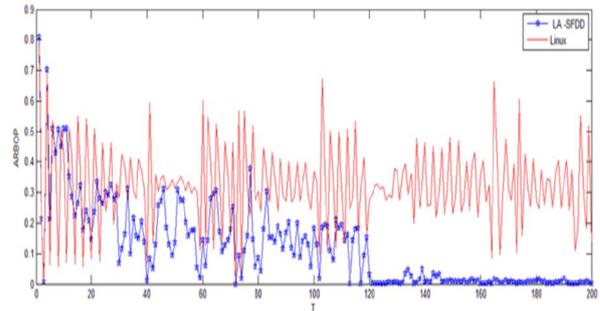

Fig. 7 buffer occupancy percentage of attack requests

Fig. 7 shows the percent occupancy of buffers by attack requests. The graph shows that the amount of buffer space allotted to attack requests are lower in LA defense algorithm, which in turn shows that a lower number of attack requests have been able to create half-open connections.

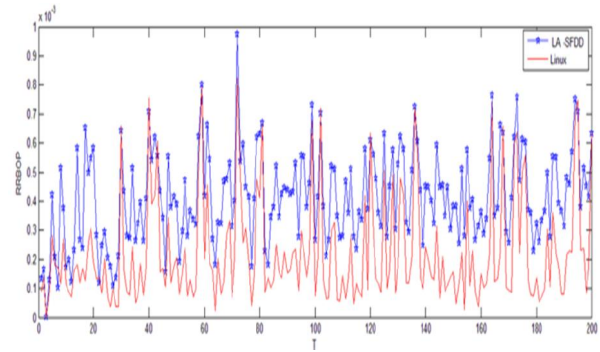

Fig. 8 buffer occupancy percentage of regular requests

Fig. 8 shows percent occupancy of buffer by normal requests. The LA defense algorithm method allocates a higher present of buffer space to itself, and thus increases level of service to privileged users.

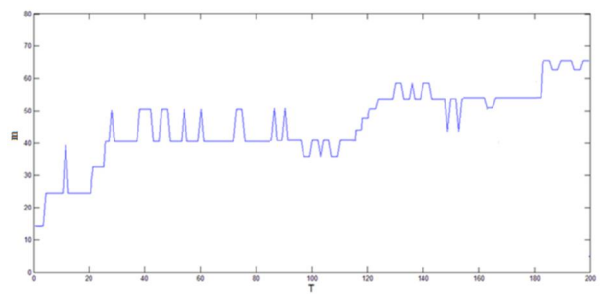

Fig. 9 *m* changes

Fig. 9 shows the graph of variations in m.

Figure 6 shows some simulation results to study about Ploss of the proposed defense. It presents Ploss for a wide range of attack intensity from $k= [0, 2]$ in which Ploss decreases as attack intensity grows up. But, it can be seen in Figure 4, 5 and 6 that when $h$ and $m$ is tuned dynamically by LA, Ploss is remarkably lower than the cases in which $h$ and $m$ are set statically to 75 and 128, respectively. Figure 4,5 shows these dynamic values of $h$ and $m$ generated by LA algorithm in our defense scheme.

Figure 8 shows how the proposed defense scheme improves performance of the under attack server in term of the Pr. We can observe in Figures 7 that the proposed algorithm, keeps Pa in lower level comparing to the case that uses fix values for $h$ and $m$. These better results are coming from dynamic and intelligent setting of $h$ and $m$ shown in Figure 4, 5. While Linux uses fixed value for $h$ this figure shows that when $k$ increases and attack intensity goes up, LA decreases $h$. Hence, long life half-open connections that typically are attack connection are closed and this makes new capacity to accept new legal connections. On the other hand, in contrast with Linux that uses fixed values for $m$, Figure 9 shows that when $k$ increases LA increases $m$. This makes new capacity for coming legal connections and hence decreases rejection probability of legal connections.

## 6. Conclusion

This paper represented a novel approach for defense against SYN-flooding attacks. We used a simple queuing model, to show important metrics of a network under DoS attacks. Then, we mapped the problem of SYN-flooding attack as an optimization problem and then employed LA technique to solve this problem. We tuned holding time and maximum allowable number of half-open connections parameters, dynamically to achieve high performance for the dynamic conditions of the network by LA technique. Simulation results confirmed success of the proposed approach.

**Future Works**

We offer the following suggestions for developing defense mechanisms to confront the SYN-flooding attacks. One may use the method of seek/search and filtering of the packets and integrate this with some of the working algorithms. By filtering the packets when they enter the server, the system will be more defensive to the attacks. The Genetic Algorithms, Combined Learning Automata Algorithm, the PSO, the Cellular Learning Algorithm, and the SWARM Robotic can be good future works to optimize the defense methods against the DoS and Distributed DoS attacks.

**Acknowledgments**

This study was carried out as a part of research project in Islamic Azad University, Ardabil Branch. We are grateful to the academic Colleagues and all persons who helped us during different phases of the study. The authors wish to thank Islamic Azad University, Ardabil Branch for supporting our research projects and financial support for this investigation.

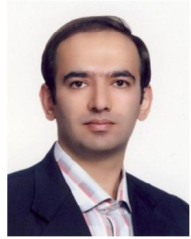

**Masoud Bekravi** received his B.Sc. and M.Sc. degrees in Electrical Engineering and Computer Engineering in 2004 (Islamic Azad university-Ardabil branch) and 2007 (Islamic Azad university-Arak branch) respectively. Now he is faculty member in the CE department, Islamic Azad University Ardabil branch. His research interests are power system, artificial intelligence, swarm robotics and Computer systems.

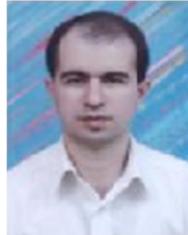

**Shahram Jamali** is an assistant professor leading the Autonomic Networking Group at the Department of Engineering, University of Mohaghegh Ardabili. He teaches on computer networks, network security, computer architecture and computer systems performance evaluation. Dr. Jamali received his M.Sc. and Ph.D. degree from the Dept. of Computer Engineering, Iran University of Science and Technology in 2001 and 2007, respectively. Since 2008, he is with Department of Computer Engineering, University of Mohaghegh Ardabil and has published more than 50 conference and journal papers.

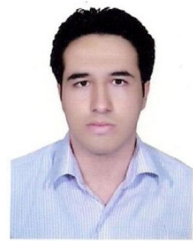

**Gholam Shaker** received his B.Sc. from Payam Noor University Ardabil, Iran, a M.Sc. (2011) from IAU (Islamic Azad University Zanjan branch) University. All are in computer engineering. he is currently Teaching in Department of Computer Science at IAU University. His research interests include Quality of Service and IDS.